\documentclass{article}
\usepackage{smc2019}
\usepackage{times}
\usepackage{ifpdf}
\usepackage[english]{babel}
\usepackage{cite}
\usepackage{color}
\usepackage{caption}
\usepackage{subcaption}
\usepackage{multirow}
\usepackage{float}

\def\papertitle{A Framework for Multi-f$_0$ Modeling in SATB Choir Recordings}
\def\firstauthor{Helena Cuesta}
\def\secondauthor{Emilia G\'omez}
\def\thirdauthor{Pritish Chandna}

\newif\ifpdf
\ifx\pdfoutput\relax
\else
   \ifcase\pdfoutput
      \pdffalse
   \else
      \pdftrue
\fi

\ifpdf 
  \usepackage[pdftex,
    pdftitle={\papertitle},
    pdfauthor={\firstauthor, \secondauthor, \thirdauthor},
    bookmarksnumbered, 
    pdfstartview=XYZ 
   ]{hyperref}

  \usepackage[pdftex]{graphicx}
  \graphicspath{{./figures/}}
  \DeclareGraphicsExtensions{.pdf,.jpeg,.png}

  \usepackage[figure,table]{hypcap}

\else 
  \usepackage[dvips,
    bookmarksnumbered, 
    pdfstartview=XYZ 
  ]{hyperref}  

  \usepackage[dvips]{epsfig,graphicx}
  \graphicspath{{./figures/}}
  \DeclareGraphicsExtensions{.eps}

  \usepackage[figure,table]{hypcap}
\fi

\hypersetup{
    colorlinks,%
    citecolor=black,%
    filecolor=black,%
    linkcolor=black,%
    urlcolor=black
}

\title{\papertitle}

 \threeauthors
   {\firstauthor} {Music Technology Group\\ 
   Universitat Pompeu Fabra \\ %
     {\tt \href{mailto:helena.cuesta@upf.edu}{helena.cuesta@upf.edu}}}
   {\secondauthor} {Universitat Pompeu Fabra and\\ 
Joint Research Centre (EC) \\ %
     {\tt \href{mailto:emilia.gomez@upf.edu}{emilia.gomez@upf.edu}}}
   {\thirdauthor} {Music Technology Group\\ 
   Universitat Pompeu Fabra \\ %
     {\tt \href{mailto:pritish.chandna@upf.edu}{pritish.chandna@upf.edu}}}

\begin{document}
\capstartfalse
\maketitle
\capstarttrue
\begin{abstract}
Fundamental frequency (f$_0$) modeling is an important but relatively unexplored aspect of choir singing. Performance evaluation as well as auditory analysis of singing, whether individually or in a choir, often depend on extracting $f_0$ contours for the singing voice. However, due to the large number of singers, singing at a similar frequency range, extracting the exact individual pitch contours from choir recordings is a challenging task. In this paper, we address this task and develop a methodology for modeling pitch contours of SATB choir recordings. 
A typical SATB choir consists of four parts, each covering a distinct range of pitches and often with multiple singers each. We first evaluate some state-of-the-art multi-f$_0$ estimation systems for the particular case of choirs with a single singer per part, and observe that the pitch of individual singers can be estimated to a relatively high degree of accuracy. We observe, however, that the scenario of multiple singers for each choir part (i.e. unison singing) is far more challenging. In this work we propose a methodology based on combining a multi-f$_0$ estimation methodology based on deep learning followed by a set of traditional DSP techniques to model f$_0$ and its dispersion instead of a single f$_0$ trajectory for each choir part. We present and discuss our observations and test our framework with different singer configurations. 

\end{abstract}

\section{Introduction}\label{sec:introduction}
Singing in a SATB (Soprano, Alto, Tenor, Bass) choir is a long standing and well enjoyed practice, with many choirs following this format across different languages and cultures. Performances are based on scores, which provide linguistic, timing and pitch information for the singers in the choir to follow. Professional choirs practice for years to ensure that their performance is \textit{in tune} with a reference pitch; however, due to the mechanism of voice production and expressive characteristics, the pitch of the individual voices in the choir often deviates from the theoretical pitch as indicated in the score. As a consequence, analysis and evaluation of a choir performance depends on the combination of pitches produced by individual singers in the choir. Through history, conductors, teachers, and critics have relied on their own interpretation of pitch and harmony, while listening and/or evaluating a choir. In recent years, a few automatic analysis and evaluation systems have been proposed \cite{cuesta2018icmpc,Dai2017} to provide an informed analysis of choirs in terms of intonation. In general, these systems require the extraction of accurate pitch contours for individual vocal tracks, which has hitherto been a roadblock for analysis, as multi-f$_0$ extraction systems are not able to provide sufficient pitch precision and accuracy to drive analysis systems from full mixed choir recordings. This can primarily be pinned down to the fact that in a choral recording, multiple singers with similar timbres are singing in harmony, and even the same notes within each choir section, leading to overlapping harmonics, which are difficult to isolate. While several multi-f$_0$ estimation systems have been designed for music with easily distinguishable sources, e.g. music with vocals, guitar, bass and drums, very few research has been carried out in the domain of vocal ensembles, be it because of the lack of annotated datasets or because modeling several people singing very close frequencies, i.e. in unison, is very challenging in terms of f$_0$ resolution.

In this work we address the computational modeling of pitch in choir recordings. In order to do that, we first evaluate how a set of multi-f$_0$ estimation algorithms perform with vocal quartets and try to identify their main limitations. Then, we use the evaluation results to select the best-performing algorithm and use it to extract a first approximation of the f$_0$ of each choir section. In the second step we use a set of traditional DSP techniques to increase the pitch resolution around the estimated f$_0$s and model f$_0$ dispersion. The main focus of this adaptation is not to obtain an accurate f$_0$ estimate for each voice inside each choir section, but to model the distribution of f$_0$ of a choir section singing in unison, measured through the dispersion of pitch values across each part.

The rest of the paper is organized as follows: Section \ref{sec:sota} provides a brief overview of the current state-of-the-art for multi-f$_0$ extraction. Section \ref{sec:problemdef} describes the limitations of current systems to characterize f$_0$ in unison performances. Then, in Section \ref{sec:eval} we define the evaluation metrics commonly used in the field, followed by Section \ref{sec:dataset} presenting the dataset used in this study. Section \ref{sec:initresults} discusses the initial evaluation of state-of-the-art methodologies on our particular material. Following this, Section \ref{sec:refinement} presents a novel approach to model unison recordings by combining a multi-f$_0$ estimation algorithm with traditional DSP techniques. Section \ref{sec:results} presents and discusses the results and limitations of the proposed system , and finally in Section \ref{sec:conc} we provide some conclusions on the method and comments on future research that we intend to carry out. 

\section{State Of The Art}
\label{sec:sota}
Multi-f$_0$ estimation involves the detection of multiple concurrent $f_0$ from an audio recording \cite{McLeod2017} and it is a core step of the task of automatic music transcription (ATM): converting an acoustic musical signal into some form of musical notation \cite{benetos2013atm}. We briefly summarize a set of multi-f$_0$ estimation methods that can be applied to vocal music, although they try to estimate f$_0$ values of individual singers, while we address the modeling of f$_0$ of unison singing.

Duan et al.\cite{duan2010multiple} presented an approach to multi-f$_0$ estimation using maximum-likelihood, where they model two different regions of the input power spectrum: the peak region, comprising the set of frequencies that are within a distance \textit{d} of the peak frequencies, and the non-peak region, which is the complement of the peak region. The input signal is normalized and the power spectrum is computed frame-wise. A set of spectral peaks are extracted using a peak detection algorithm, and then several f$_0$ candidates are computed in the range of one semitone around each peak. For each time frame, the f$_0$ of each source is estimated by maximizing the probability of having harmonics that explain the observed peaks and minimizing the probability of having harmonics in the region where no peaks were observed. This is accomplished optimizing the parameters of a likelihood function that combines the peak region likelihood and the non-peak region likelihood, treated as independent sets. They use monophonic and polyphonic training data to learn the model parameters. Their system also estimates polyphony, i.e. how many sources there are in the mix, which define the number of f$_0$ the model should estimate at each frame. Finally, a post-processing step using information for neighbouring frames is implemented to make the pitch predictions more stable. This process of refining the f$_0$ estimates, however, removes duplicate estimates, which a problem in the case of several sources producing the same f$_0$, i.e. unison singing. The system parameters were learned using training data consisting of mixes of individual monophonic note recordings from 16 instruments including flute, saxophone, oboe, violin, bass, and violin among others. Then, they evaluated the algorithm on 4-part Bach chorales performed by a quartet: violin, clarinet, tenor saxophone and bassoon. These details about the data they used to train and evaluate the system are very relevant for our research, since given the lack of vocal data in the training and evaluation stages, we expect the system to perform worse in choral music.

Another relevant system for multiple f$_0$ estimation is the one developed by Klapuri \cite{klapuri2006multiple}, which estimates each f$_0$ in the mixture iteratively: at every step, the system detects the most predominant f$_0$ and its corresponding harmonics are then substracted from the spectrum. In this case, the input signal is passed through a bank of linear band-pass filters that resemble the inner ear behaviour in terms of frequency selectivity. Then, the output signal at each band is processed in a nonlinear manner to approximate the firing activity in the auditory nerve. After this pre-processing steps, a frequency-domain signal representation is obtained by combining the band spectra, which is then used to extract a pitch salience function to emphasize the fundamental frequencies present in the signal. From this representation, multiple f$_0$ values are estimated iteratively: at each step, a $f_0$ value is estimated and its harmonic partials are removed from the spectrum. This step is repeated until a $f_0$ value is estimated for each source. In \cite{klapuri2006multiple} he also implements a polyphony estimator, which determines how many $f_0$ values need to be extracted and therefore the number of iterations. The system was evaluated with a collection of mixtures of individual sounds (some of them from the same source as in \cite{duan2010multiple}). The authors does not explicitly mention any vocals in the dataset, and therefore we assume the method is not optimized for our particular material.

Schramm and Benetos \cite{schramm2017automatic} presented a method specifically designed for multi-f$_0$ estimation in \textit{a cappella} vocal ensembles. They use a two-step approach: first, a system based on probabilistic latent component analysis (PLCA) employs a fixed dictionary of spectral templates to extract the first frequency estimates; as a second step, a binary random forest classifier is used to refine the f$_0$ estimates based on the overtones properties. Spectral templates are extracted from recordings of multiple singers singing pure vowels in English. These recordings belong to the RWC dataset \cite{goto2003rwc}. This method uses the normalized variable-Q transform (VQT) as input, which is then factorized using the expectation-maximization algorithm to estimate the parameters of the model. As opposed to the previously presented methods, this one is focused on vocal ensembles and it is trained and evaluated with such data, i.e. vocal quartets, one singer per part. They use a f$_0$ resolution of 20 cents, which is enough for transcription purposes but not to deal with frequencies as close to each other as in unisons.

Another method for multi-f$_0$ estimation designed for the case of multiple singers is the one presented by Su et al. \cite{Su2016}. The authors claim that data is crucial to develop and evaluate such systems, and yet there is not a labeled multi-f$_0$ dataset for choir, which is one of the most common type of music through the ages and cultures. Their work has two separate parts: first, they present a novel annotated dataset of choir and symphonic music; then, they build an unsupervised approach to multi-f$_0$ estimation using advanced time-frequency (TF) analysis techniques such as the concentration of time and frequency method. According to their paper, these techniques help improving the stabilization of pitch, which is interpreted in three dimensions: frequency, periodicity, and harmonicity. 

Recent advancements in deep learning based systems have led novel deep learning based multi-f$_0$ extraction systems, designed to be agnostic to the exact source of the pitched instruments in the mix. \textit{DeepSalience} \cite{bittner2017deep} is one of the most recent systems for multi-instrument pop/rock songs and mixtures. The model leverages harmonic information provided by a HCQT transform, comprising $6$ constant-Q transforms (CQT), with a convolutional neural network (CNN) to extract pitch salience from an input audio mixture. The network is fully convolutional with 5 convolutional layers, it uses batch normalization and rectified linear units (ReLU) at each output. The final layer of the network uses logistic activation, mapping each bin of the output to the range [0,1], representing pitch salience. It is trained using cross-entropy minimization. This pitch salience essentially predicts the probabilities of the underlying pitches being present in the input signal with a resolution of 20 cents. Then, using this salience intermediate representation, they use a threshold to estimate multiple frequencies at each frame. 

These methods are capable of extracting multiple f$_0$ from a great variety of audio signals, including music and speech. Most of them are designed for polyphonic signals where each melody is produced by a single source: one instrument or singer. However, the subject of our study are choirs, which involve unison ensemble singing, i.e. performances where several people sing the same notes. Unison recordings are challenging for multi-f$_0$ estimation because of the possible imprecision in the pitch produced by multiple singers or musicians \cite{Su2016}. Since we focus on the analysis and synthesis of choral singing, it is crucial to take this aspect into account to build models that consider these pitch imprecision.
\section{Problem Definition and Approach}\label{sec:problemdef}
The characterization of pitch distribution in unison and choir singing has not been widely studied. Most of the research in this topic is authored by Sundberg \cite{sundberg1987science} and Ternstr\"om, who published a review on choir acoustics \cite{Ternstrom2002} and carried out several experiments to study f$_0$ dispersion in unison singing \cite{Ternstrom1991}. The authors define f$_0$ dispersion as the small deviations in f$_0$ between singers that produce the same notes in a unison performance. This magnitude is directly related to the \textit{degree of unison}, which Sundberg defines as the agreement between all the voices sources. In a later work, Jers and Ternstr\"om \cite{jers2005intonation} measured the dispersion between singers and found it to range between 25 and 30 cents. 

In multi-f$_0$ estimation systems, we usually focus on the extraction of a single pitch per source, and state-of-the-art algorithms would then provide a f$_0$ value for each choir section. However, several singers produce slightly different values in each of the voices of a choir. Then, the question of which is the correct value to be estimated  arises: most multi-f$_0$ estimation algorithms do not have enough resolution to discern the individual pitches, which leads to a potentially imprecise estimation. This suggests that unison performances need to be treated in a different way. Ternstr\"om \cite{Ternstrom1991} claims that while solo singing has tones with well-defined properties, i.e. pitch, loudness, timbre, unison ensemble singing has tones with statistical distributions of these properties, and we need to consider those when modeling them. 

\begin{center}
\begin{figure}[h]
    \centering
    \includegraphics[width=\columnwidth]{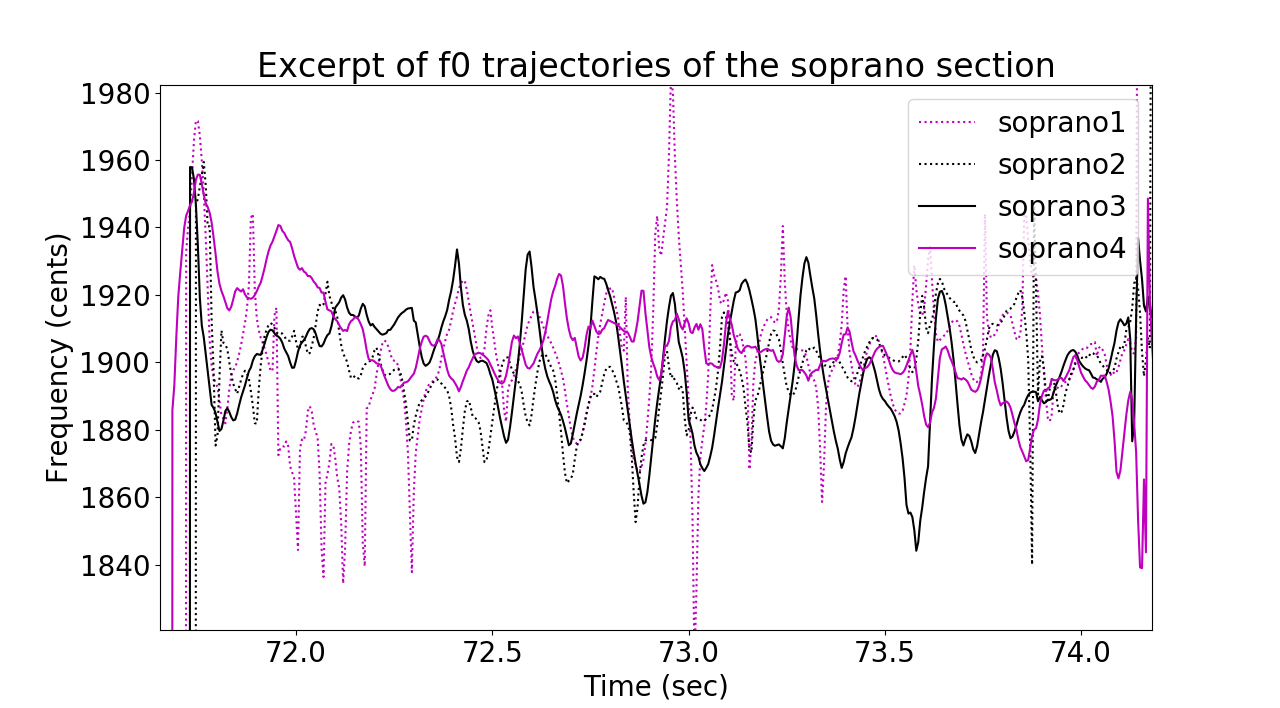}
    \caption{F$_0$ curves of four sopranos singing the same note. We see how the curves oscillate and differ from each other.}
    \label{fig:sopranos}
\end{figure}
\end{center}

In a recent study, Cuesta and al. \cite{cuesta2018icmpc} created the Choral Singing Dataset (see Section \ref{sec:dataset}) to analyze f$_0$ dispersion in unison ensemble singing by modeling the distribution of fundamental frequencies. Using individual tracks for each singer, they extracted f$_0$ curves and computed the mean f$_0$ as the perceived pitch and the standard deviation of the distribution as the f$_0$ dispersion. In Figure \ref{fig:sopranos} we display an example of the f$_0$ trajectories of four sopranos, where we observe that there are slight f$_0$ differences between them. This study found dispersion values ranging from 20 to 30 cents on average, depending on the choir section and the song, which agrees to previous literature \cite{Ternstrom1991}. However, this type of analysis requires an individual audio track for each singer in the mixture, and this data is difficult to obtain given that choirs are not recorded using this set up. This particular limitation leads us to explore in the present study ways of analyzing choir recordings directly from the singer mixture, which involves dealing with four different melodies (SATB), each of them involving a unison.

In this study, we propose a methodology for pitch content analysis on unison ensemble singing that has two main stages:
\begin{enumerate}
    \item{\bf{Multi-f$_0$} estimation}. In the first stage, we perform multi-f$_0$ estimation in the audio mixture in order to roughly estimate the pitches of the four voices of the choir. For this part, we evaluate the performance of a set of existing multi-f$_0$ estimation systems in the context of vocal quartets, where we have precise f$_0$ ground truth information, to select the one with a better performance.
    \item{\bf{F$_0$-dispersion modeling}} In the second stage, we refine the frequency analysis around those pitches to further characterize f$_0$ dispersion in each of the unison voices. In order to do so, we consider a DSP-based approach and adapt a method with higher frequency resolution to model each melodic source as a distribution of f$_0$s instead of a single value.
\end{enumerate} 
\section{Evaluation Methodology}
\label{sec:eval}
As mentioned in previous sections, we evaluate the performance of three state-of-the-art algorithms for multi-f$_0$ estimation in vocal quartets, e.g. SATB with a single singer per section, in order to investigate which method is more suitable for this music material. We consider the methods proposed by Klapuri (KL) \cite{klapuri2006multiple}, Schramm et al. (SCH) \cite{schramm2017automatic} and Bittner et al. (DS) \cite{bittner2017deep}, all of them publicly available and representative of the state of the art in the area.
\subsection{Dataset}
\label{sec:dataset}
There are very few datasets of choral music which are annotated in terms of f$_0$. In our experiments, we take advantage of the Choral Singing Dataset\footnote{Choral Singing Dataset: \url{https://zenodo.org/record/1319597}} further described in \cite{cuesta2018icmpc}. This dataset was recorded in a professional studio and contains individual tracks for each of the 16 singers of a SATB choir, i.e. 4 singers per choir section. Although each section was recorded separately, synchronization between all audio tracks was achieved using a piano reference and a video of the conductor that singers followed during the recording.

This dataset comprises three different choral pieces: \textit{Locus Iste}, written by Anton Bruckner, \textit{Ni\~no Dios d'Amor Herido}, written by Francisco Guerrero, and \textit{El Rossinyol}, a Catalan popular song; all of them were written for 4-part mixed choir. This dataset is more suitable for this study than the one presented in \cite{Su2016}: having the individual tracks of each singer allows us to create \textit{artificial} mixes between voices, e.g. vocal duets or quartets, small choir, large choir...etc. Using different combinations of all 16 singers, we created 256 SATB quartets for each piece, which represent all  possible combinations of singers taking into account the voice type restriction, i.e. we need one singer per voice. These vocal quartets are used to evaluate the performance of the three algorithms.
%
\subsection{Multi-f$_0$ evaluation metrics}\label{sec:metrics}
In multi-f$_0$ estimation systems, there are multiple f$_0$ values per frame \textit{n}. Following the terminology used by Bittner \cite{bittner2018phd}, we define the ground truth value(s) in frame $n$ as $f$[n] and the estimation as $\hat{f}$[n], which denote the pitches of all active sources in that frame.

For a given frame $n$ we denote as true positives, TP[$n$], the number of correctly transcribed pitches, and as false positives, FP[$n$], the number of pitches present in the estimation, $\hat{f}$[$n$], which are not present in the ground truth, $f$[$n$]. Similarly, the false negatives value, FN[$n$], measures the number of pitches present in the ground truth which are not present in the estimation. Based on these, we define the following set of metrics: \textit{accuracy, precision and recall}, and a set of errors: the substitution error ($E_{sub}$), miss error ($E_{miss}$), and false alarm error ($E_{fa}$). Finally, total error, $E_{tot}$ is reported as the combination of $E_{sub},E_{miss}$ and $E_{fa}$. 

All the presented evaluation metrics also have their associated \textit{chroma} versions, which considers an estimated f$_0$ to be correct if it is one octave apart from the corresponding target pitch. For more details about these metrics we refer the reader to \cite{bittner2018phd}.
\section{Multi-f$_0$ Estimation Results}
\label{sec:initresults}
All the SATB quartets of the dataset were evaluated in terms of multi-f$_0$ estimation: by means of the individual tracks, we extracted f$_0$ curves for every singer using the spectral-amplitude autocorrelation (SAC) method \cite{sac2013} and we then combined them to create the multiple f$_0$ ground truth at each frame. 

\begin{center}
\begin{figure}[h]
    \centering
    \includegraphics[width=\columnwidth]{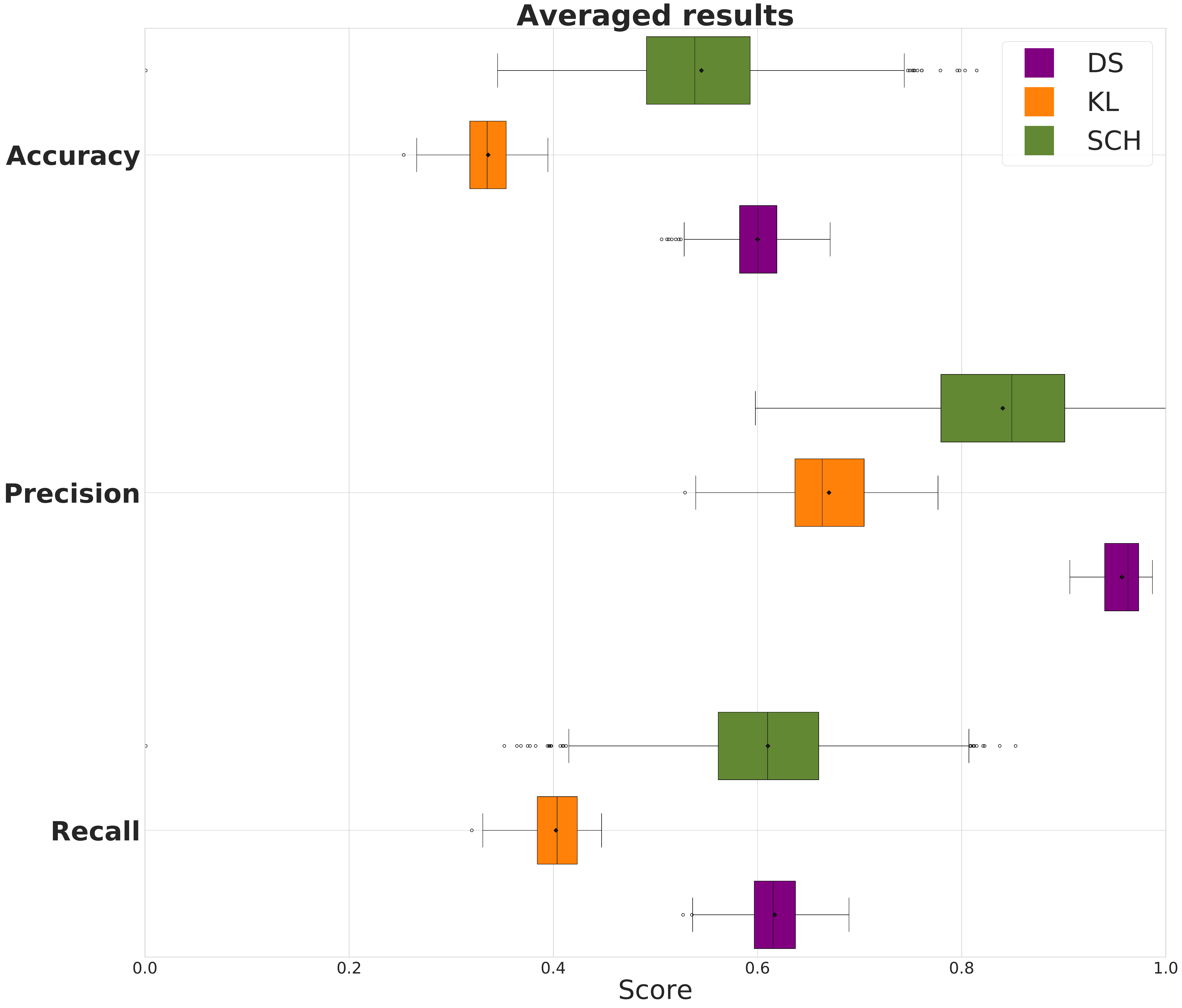}
    \caption{Accuracy, precision and recall for each of the three algorithms (DS, DU, SCH) averaged over all the dataset, i.e. all the SATB quartets.}
    \label{fig:results_quartet}
\end{figure}
\end{center}

A summary of the results is displayed in Figure \ref{fig:results_quartet}, where we present the accuracy, recall and precision averaged for each of the algorithms in the three songs of the dataset. We observe that \textit{DeepSalience} (DS) outperforms Klapuri (KL) and Schramm (SCH). It is also interesting to point out that the difference between these metrics and their \textit{chroma} versions is very small, thus suggesting that the three algorithms are fairly robust in terms of octave errors. We also observe that the algorithm by Schramm et al. has a higher variability with respect to the other ones, suggesting that its performance is highly dependent on the input signal. Also, it is important to mention that while KL and DS predict multi-f$_0$ values from a long audio file, i.e. a full choral recording, we splitted our audio material in shorter clips (each of them 10 seconds long) to evaluate SCH method: the PLCA algorithm employed in this method is computationally very expensive and we could not obtain results using the full recordings. 
In terms of error analysis, Table \ref{tab:errorresults} provides the average errors for each of the algorithms. These results suggest that extracting multiple frequencies from a vocal ensemble is a very challenging task, since the total error is almost 40\% in the best performing method. A part from this, we can extract a few more insights: all algorithms have a very low false alarm error, which means that they almost never report an f$_0$ when there is not one in the ground truth; in addition, \textit{DeepSalience} does a good job regarding the substitution error, which means that it rarely reports a wrong f$_0$. However, the miss error is pretty high, especially in Klapuri's algorithm, which means that there are a lot of f$_0$ that are not extracted. In the case of Schramm's method, though, the miss error is lower, suggesting that their voice assignment step improves the performance of the multi-f$_0$ estimation. We could also relate these differences to the fact that SCH is the only method designed for and trained with singing voice data. However, given the length limitation of SCH and based on the overall results, we select \textit{DeepSalience} as the method to be used in the first stage of our model. 
\begin{table}[t]
 \begin{center}
 \begin{tabular}{|l||l|l|l|}
  \hline
  & \textbf{DS} & \textbf{SCH} & \textbf{KL} \\ 
  \hline\hline
  Substitution error & 2.3\% & 10\% & 12\% \\ 
  \hline
  Miss error & 35\% & 28\% & 48\% \\ 
  \hline
  False alarm error & 0.4\% & 1.5\% & 8\% \\ 
  \hline
  Total error & 38\% & 40\% & 67\% \\ 
  \hline
 \end{tabular}
\end{center}
 \caption{Summary of error metrics in multi-f$_0$ estimation .}
 \label{tab:errorresults}
\end{table}
\section{F$_0$ Dispersion Modeling}\label{sec:refinement}
The first step of our method presented above uses \textit{DeepSalience} to extract multiple pitch estimations at each frame of the audio input. In the ideal case, at this stage we would obtain one f$_0$ value for each choir section; however, as discussed in Section \ref{sec:initresults}, although \textit{DeepSalience} is the best-performing algorithm from the evaluated set, there are still some errors in the output.

\begin{figure}
\centering
\begin{subfigure}{\columnwidth}
\includegraphics[width=\columnwidth]{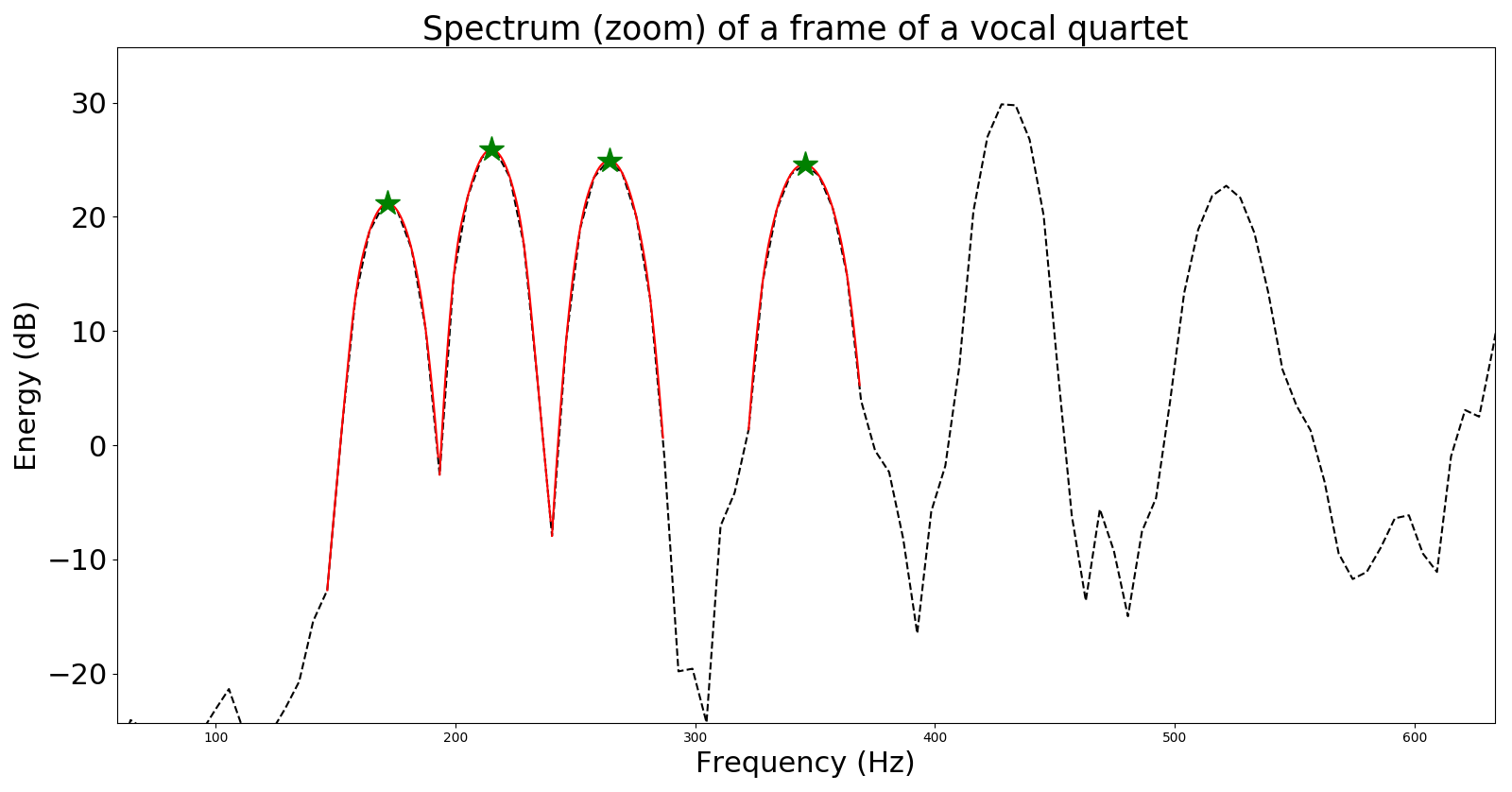}
\end{subfigure}
\hspace{01em}
\begin{subfigure}{\columnwidth}
\includegraphics[width=\columnwidth]{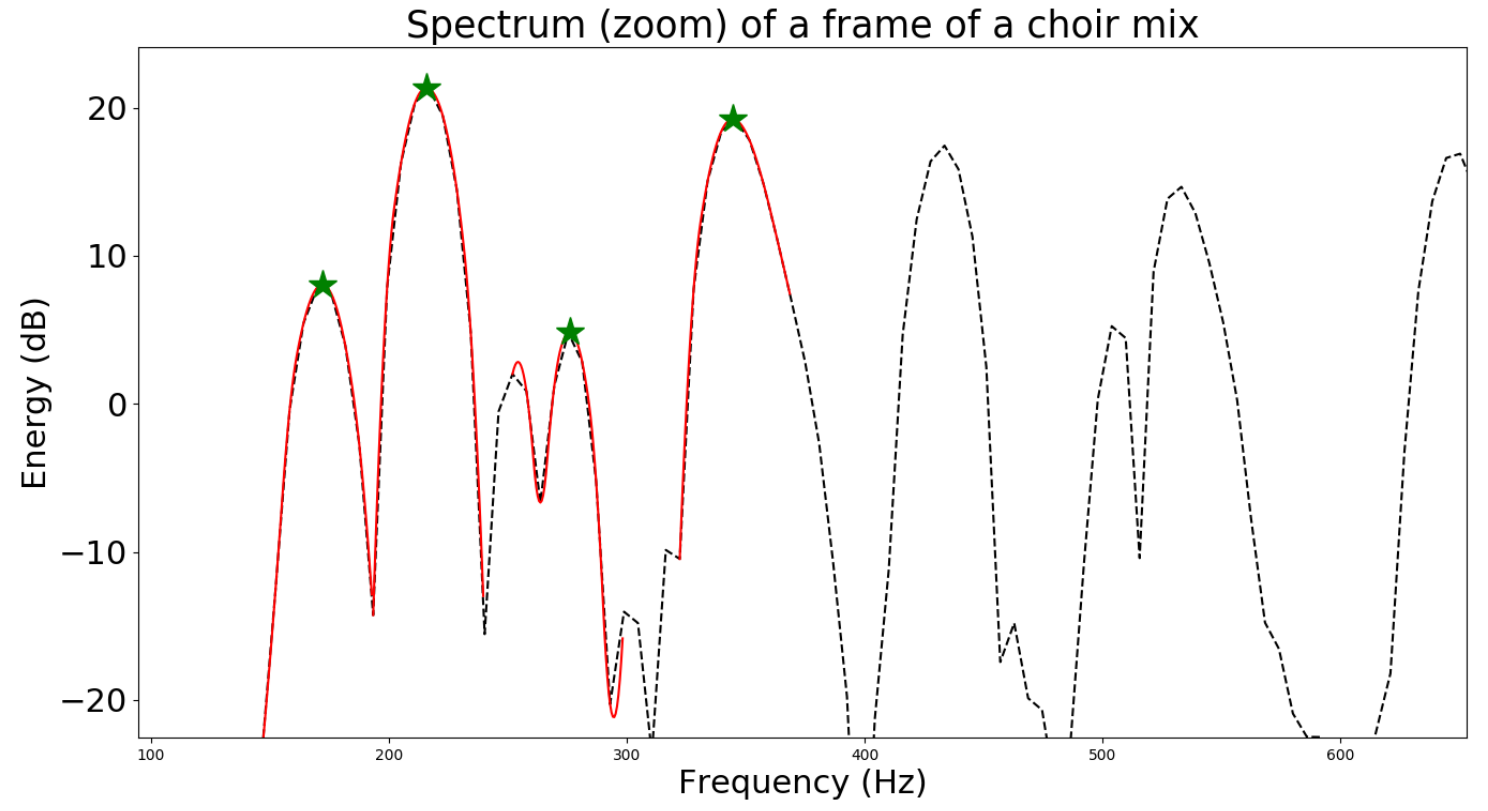}
\end{subfigure}
   \caption{Example frames of the spectrum with the interpolated peaks corresponding to the estimated f$_0$s (green asterisk). The top plot corresponds to a 4-singers audio input (quartet), while the bottom plot corresponds to the 16-singer audio input (full choir).}
   \label{fig:peakint} 
 \end{figure}

\begin{center}
\begin{figure}
    \centering
    \includegraphics[width=\columnwidth]{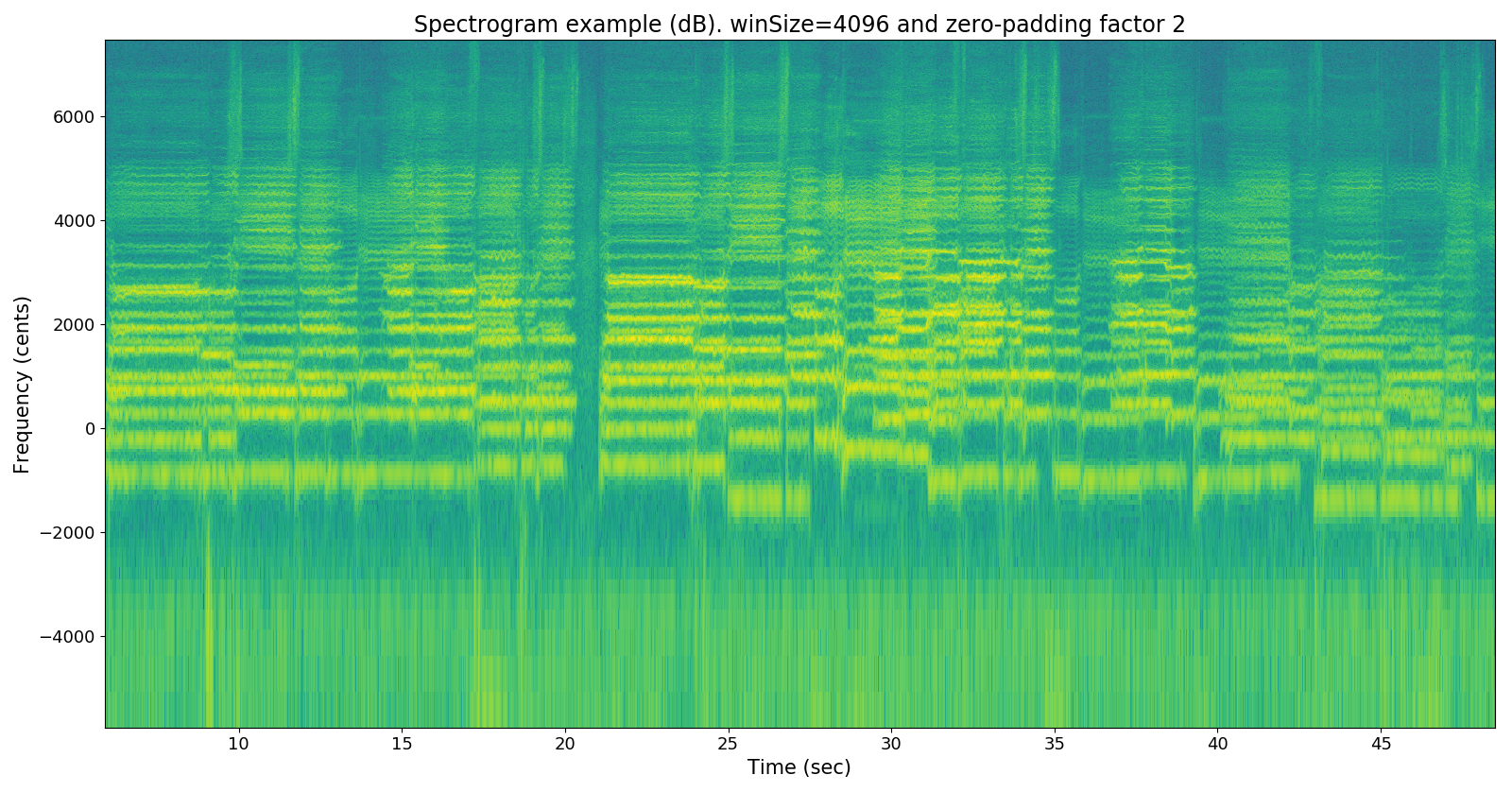}
    \caption{Spectrogram (zoomed) of the piece Locus Iste computed using a Hanning window and $N=4096$ zero-padded to twice its length, resulting in an FFT size of 8192.}
    \label{fig:spec}
\end{figure}
\end{center}

In the second stage of our method we consider traditional DSP techniques to increase the frequency resolution of our model. We compute the spectrogram of the input audio signal using a Hanning window of 4096 points zero-padded to twice its length, resulting in an FFT size of 8192. An excerpt of this spectrogram is displayed in Figure \ref{fig:spec} with magnitude in dB and the frequency axis in cents, where we observe that f$_0$ values for each choir section are well-separated.  \\ 
With this time-frequency representation, we then locate each of the estimated fundamental frequencies (\textit{DeepSalience} output), which will ideally match one of the spectral peaks. Even though we use a large FFT size, since we want to obtain a high pitch resolution, we interpolate the peaks and recompute the peak locations as the maximum value of each interpolated peak.  This process is illustrated in Figure \ref{fig:peakint}, where the top and bottom plots correspond to a vocal quartet and full choir spectrum, respectively. The dashed black line represents the original spectrum, while the red solid lines and the green asterisk correspond to the interpolated peaks. We observe that the peaks in the full choir case (bottom) have less energy and are a bit more noisy than the vocal quartet ones (see third and sixth peaks for example).
 
Once we have this information, we compute the bandwidth of each peak as a measure of the dispersion of the f$_0$ distribution in the unison case. Remember that our aim is to characterize the distribution of f$_0$ for each choir section rather that obtaining a single f$_0$ value. For each choir section, we find and interpolate the peak in the spectrum and consider the peak frequency as the mean frequency of the distribution and its bandwidth as its dispersion. The bandwidth is expressed in cents (computed with a reference frequency of 220 Hz) and computed as follows:
\begin{equation}\label{eq:bw}
    f_{0_{dispersion}} = b_2 - b_1
\end{equation}
where $b_2$ and $b_1$ are the frequency bins around the spectral peak where the amplitude of the spectrum decays $3$ dB. Note that in this first approach we do not take into account the window type and size used in the spectral analysis, although they influence the peak bandwidth. In further studies, we plan to study and document the effect of these two analysis parameters in the dispersion computation.
%
 %
\section{Results}\label{sec:results}
\begin{center}
\begin{figure}
    \centering
    \includegraphics[width=\columnwidth]{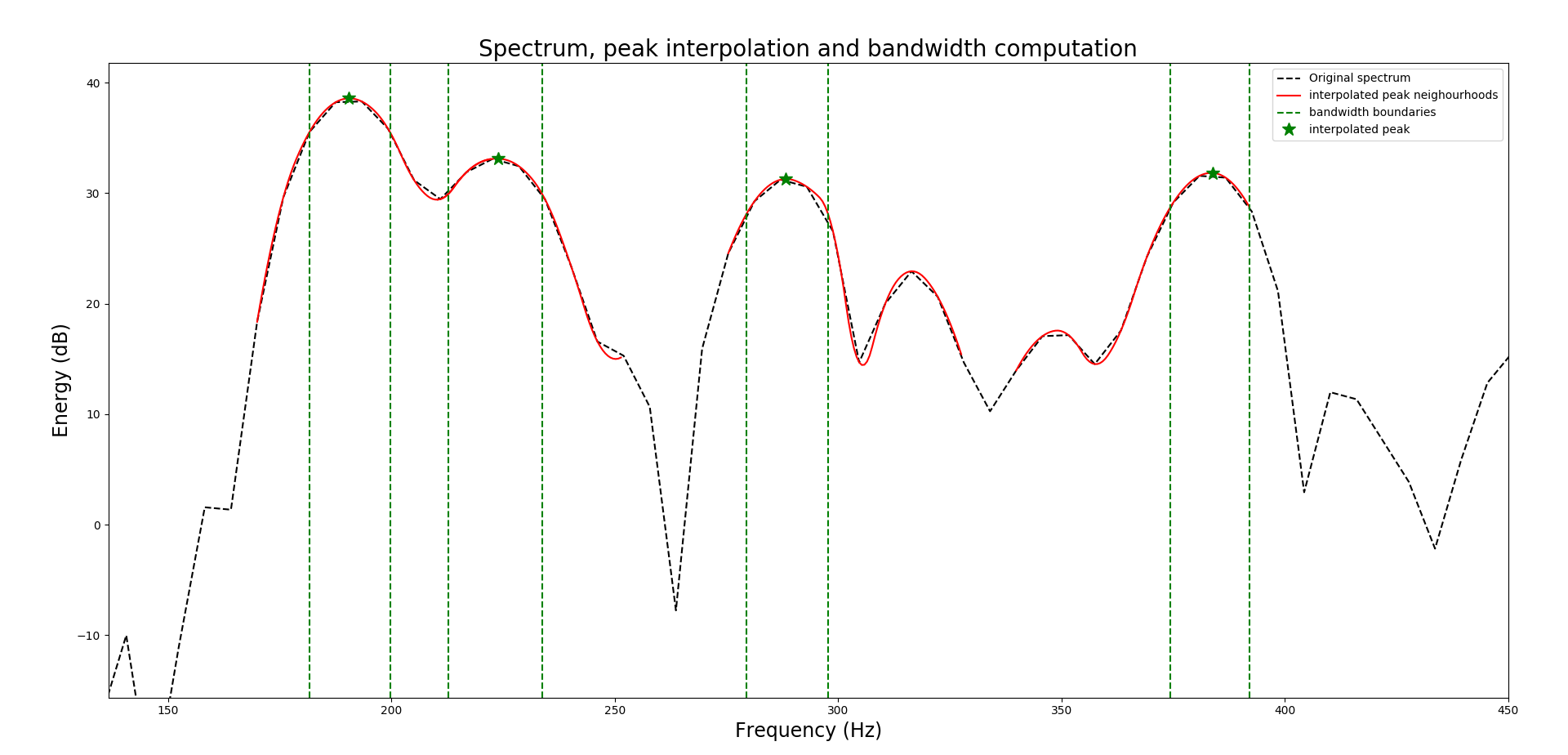}
    \caption{Example frame of the spectrum with the interpolated peaks and the boundaries for the bandwidth computation.}
    \label{fig:examplebw}
\end{figure}
\end{center}

In this Section we present the results of the dispersion analysis averaged by piece and by choir section. Since the first step of the presented framework outputs noisy multi-f$_0$ estimations, the results of the second step displayed here are obtained using the ground truth pitch values to locate the peaks. This allows us to perform a better evaluation of the dispersion computation method, since the errors that come from the multi-f$_0$ estimation are not present here. In the real application, the ground truth pitches would be replaced by the output of the selected multi-f$_0$ estimation algorithm. Figure \ref{fig:examplebw} shows another example frame of the analysis, with the peak interpolation and also vertical lines showing the bandwidth delimitation.
%
%
Table \ref{tab:dispersionresults} provides the f$_0$ dispersion results averaged by choir section and piece. \textit{BTAS} stand for \textit{bass}, \textit{tenor}, \textit{alto}, \textit{soprano}, and \textit{Q} and \textit{CM} are short for \textit{quartet} and \textit{choir mix}, meaning that the dispersion values belong to a 4-singers and 16-singers setting, respectively. We would expect the dispersion values to be larger in the 16 singers case, because having several voices producing similar frequencies might generate wider spectral peaks. The differences on average are not very strong; however, we conducted an independent-samples t-test to compare the performances of vocal quartets with the full choir and found that the differences were significant. For example, there was a significant difference in the bass section of the quartet singing El Rossinyol (M=181,SD=39) and the bass section in the choir mix (M=191,SD=79), $\alpha=0.05, p<0.001$. This tendency applies to the whole dataset, although the effect size is small (around 0.2 on average) and therefore we might need a deeper analysis to find out if the differences are not only statistically significant but also relevant in terms of pitch content.

These results can not be directly compared to any ground truth, since previous studies modeled pitch dispersion in different ways, i.e. standard deviation of the distribution \cite{cuesta2018icmpc} or bandwidth of the partial tones \cite{Ternstrom1991}. Instead, we compare the tendency of the results with the ones obtained by the authors in a previous study where individual pitch tracks, and thus accurate individual f$_0$ estimations, were used \cite{cuesta2018icmpc}. In Table \ref{tab:icmpc} we display the results from the mentioned work, where the dispersion is averaged by choir section (all three pieces together) and by piece (all four choir sections together). These results suggest that the dispersion (in cents) is higher in the bass section of a choir, which is also true for our results. Following basses, the presented results have tenors, followed by altos and finally sopranos, with the lowest average dispersion. In Table \ref{tab:icmpc} we have altos and tenors with very similar average values, and sopranos also obtained the lowest dispersion values. Therefore, although the dispersion magnitude can not be compared, the trend is very similar, suggesting that the analysis is consistent.
\begin{table}[t]
 \begin{center}
 \begin{tabular}{|c|c|c||c|c||c|c|}
  \hline
  & \multicolumn{2}{|c||}{\textbf{Locus Iste}} & \multicolumn{2}{c|}{\textbf{El Rossinyol}} & \multicolumn{2}{c|}{\textbf{Ni\~no Dios}} \\
  \hline
  & Q & CM & Q &CM & Q & CH \\ \hline
  \textbf{B} & 231  & 248 & 181  & 191 & 227 & 257 \\ 
  & (57) & \textit{(130)} & \textit{(39)} & \textit{(70)} & \textit{58} & \textit{175} \\ \hline
  \textbf{T} & 136 & 140 & 132 & 136 & 143 & 149 \\ 
  & \textit{(30)} & \textit{(38)} & \textit{(26)} & \textit{(30)} & \textit{(31)} & \textit{(40)} \\  \hline
  \textbf{A} & 100 & 104 & 98 & 103 & 105 & 110 \\ 
  & \textit{(22)} & \textit{(25)} & \textit{(19)} & \textit{(23)} & \textit{(22)} & \textit{(28)} \\ \hline
  \textbf{S} & 76 & 79 & 75 & 80 & 78 & 82 \\ 
  & \textit{(23)} & \textit{(27)} & \textit{(16)} & \textit{(20)} & \textit{(20)} & \textit{(25)} \\  \hline
 \end{tabular}
\end{center}
 \caption{Average f$_0$ dispersion results computed as the bandwidth of the peaks in the whitened spectrum. Dispersion values are in \textbf{cents}. \textit{Q} refers to a SATB quartet with one singer per section and \textit{CM} refers to a SATB choir mix with 4 singers per section. Values in italics are standard deviations also in cents.}
 \label{tab:dispersionresults}
\end{table}
\begin{table}[]
    \centering
    \begin{tabular}{|c|c|c|c|}
    \hline
       \textbf{Soprano} & \textbf{Alto} & \textbf{Tenor} & \textbf{Bass}  \\ \hline
       20.16 & 22.66 & 22.22 & 26.02 \\ \hline\hline
       \multicolumn{2}{|c|}{\textbf{Locus Iste}} & \textbf{El Rossinyol} & \textbf{Ni\~no Dios} \\ \hline
       \multicolumn{2}{|c|}{19.32}  & 27.65 & 21.33 \\ \hline
    \end{tabular}
    \caption{Averaged results of pitch dispersion from \cite{cuesta2018icmpc}. All values represent dispersion in cents, computed as the standard deviation of the pitch distribution.}
    \label{tab:icmpc}
\end{table}
%
%
\\ 

After an analysis of the results, we observe that this framework has a few limitations, including that the results highly depend on the performance of the multi-f$_0$ estimation algorithm employed in the first stage. In this paper we evaluated three algorithms which are considered state-of-the-art and \textit{DeepSalience} was selected according to the evaluation we carried out, but we hypothesize that using a system specially designed for singing voice, and even for choral singing, e.g. \cite{schramm2017automatic} and \cite{Su2016}, might improve the performance of the whole method: if the f$_0$ estimates at each frame are precise, then the peaks would be modelled correctly (as happened in the dispersion evaluation), yielding more accurate f$_0$ dispersion values. However, these methods were dismissed from the final approach because of the length limitation \cite{schramm2017automatic} and because it is not publicly available \cite{Su2016}.

The proposed framework models the f$_0$ distribution of a unison using two values: the f$_0$, extracted in the first stage and refined in the second stage, and the f$_0$ dispersion. We believe these values are a good descriptor for unison performances, but a more complex model incorporating temporal evolution analysis could also be considered and used to estimate the quality of a choir or in choral synthesis applications.

\section{Conclusions}\label{sec:conc}
In this paper we presented a framework for the f$_0$ modeling in choral singing recordings that uses a two-stage methodology: first, a deep learning based multi-f$_0$ estimation system is employed to obtain one pitch value for each choir section; second, we locate these frequencies in a whitened version of the spectrum of the input signal with a higher pitch resolution. This process allows us to model each unison as a distribution of f$_0$, characterized by two values: the mean f$_0$ and the f$_0$ dispersion. The preliminary results we obtained do not show strong differences in the dispersion between a large and a small number of singers, but more data might reveal different patterns. Since we evaluated this framework as a case study, more experiments will be done in the near future, and a deeper analysis of the relationship between the time-frequency representation and the results will be carried out to make the system more effective.

\begin{acknowledgments}
This work is partially supported by the European Commission under the TROMPA project (H2020 770376) and by Spanish Ministry of Economy and Competitiveness under the CASAS project (TIN2015-70816-R). First author is supported by FI Predoctoral Grant from AGAUR (Generalitat de Catalunya). We thank Rodrigo Schramm and Andrew McLeod for the support running their code with our audio material.
\end{acknowledgments} 

\bibliography{smc2019bib}

\end{document}